\documentclass[twocolumn, preprint2]{aastex61}

\usepackage{graphicx}
\usepackage[utf8]{inputenc}
\usepackage{hyperref}
\usepackage{amsmath}
\usepackage{multirow}

\shorttitle{Information Retrieval and Recommendation System for Astronomical Observatories}
\shortauthors{Mukund et al.}

\begin{document}

\title{Information Retrieval and Recommendation System for Astronomical Observatories}
\author{Nikhil Mukund}
\affil{Inter-University Centre for Astronomy and Astrophysics (IUCAA), Post Bag 4, Ganeshkhind, Pune 411 007, India}

\author{Saurabh Thakur}
\affil{Vellore Institute of Technology, Tamil Nadu, India}

\author{Sheelu Abraham}
\affil{Inter-University Centre for Astronomy and Astrophysics (IUCAA), Post Bag 4, Ganeshkhind, Pune 411 007, India}

\author{A. K. Aniyan}
\affil{Department of Physics and Electronics, Rhodes University, Grahamstown, South Africa}
\affil{SKA South Africa, $3^{rd}$ Floor, The Park, Cape Town, South Africa}

\author{Sanjit Mitra}
\affil{Inter-University Centre for Astronomy and Astrophysics (IUCAA), Post Bag 4, Ganeshkhind, Pune 411 007, India}

\author{Ninan Sajeeth Philip}
\affil{Department of Physics, St. Thomas College, Kozhencherry, Kerala, India}

\author{Kaustubh Vaghmare}
\affil{Inter-University Centre for Astronomy and Astrophysics (IUCAA), Post Bag 4, Ganeshkhind, Pune 411 007, India}

\author{D. P. Acharjya}
\affil{Vellore Institute of Technology, Tamil Nadu, India}

\begin{abstract}
We present a machine learning based information retrieval system for astronomical observatories that tries to address user defined queries related to an instrument. In the modern instrumentation scenario where heterogeneous systems and talents are simultaneously at work, the ability to supply with the right information helps speeding up the detector maintenance operations. Enhancing the detector uptime leads to increased coincidence observation and improves the likelihood for the detection of astrophysical signals. Besides, such efforts will efficiently disseminate technical knowledge to a wider audience and will help the ongoing efforts to build upcoming detectors like the LIGO-India etc even at the design phase to foresee possible challenges. The proposed method analyses existing documented efforts at the site to intelligently group together related information to a query and to present it on-line to the user. The user in response can further go into interesting links and find already developed solutions or probable ways to address the present situation optimally. A web application that incorporates the above idea has been implemented and tested for LIGO Livingston, LIGO Hanford and Virgo observatories. 
\end{abstract}


\section{Introduction} \label{sec:intro}

Data mining in the big data framework often encounters difficulty in both extracting the relevant information from the data and in coming up with meaningful interpretations in a highly reliable fashion \citep{fan2014challenges,wu2014data,khan2014big}. In many situations, the data comes in a format which is not suitable to store in relational databases of coherent hierarchy \citep{stephens2015big}.  The methods in which the data is stored and associated with different entities also pose the challenge in mining required information from it. For example, in a gravitational wave observatory, there will be a core science dataset with plenty of meta-data on the observation and a variety of other auxiliary datasets collected from various sensors and actuators. 

Even though data mining methods like association analysis, clustering and other machine learning techniques exist, the presentation of unstructured data into these algorithms and inference generation is not a trivial task \citep{han2011data}.  Generation of insights from big data with recommendation systems which are based on learning from unstructured text data \citep{pazzani2007content} tackle these challenges at large \citep{lavalle2011big,hu2014toward}. Descriptive recommendations and information retrieval \citep{sigurbjornsson2008flickr,gretzel2004tell} have recently have recently gained popularity and have been applied to areas like travel recommendation systems \citep{gretzel2004tell} and content personalization systems \citep{liang2006personalized}. Besides commercial applications, text summarization based content recommendation \citep{hassan2009content} is an interesting area which has high level of applicability in different areas of science and research \citep{miner2012practical,2017arXiv170505840K}. Unlike conventional rank based search systems, these  do not perform topical modeling and rank topics of recurring interest \citep{zoghbi2013words}. Topical modeling is usually done for retrieving information from a single website with multiple topics. The challenge is when different topics in a single site may be weakly linked to each other \citep{cointet2010local}. While there could be already known relations among different entities, the process of data mining and better data representation can reveal the latent unforeseen linkages among different topical entities \citep{behrens2006information}.

Large science projects, especially astronomical observatories, have plenty of data  about telescope operations, scheduling, maintenance and general observational activities all logged in text form.  Over the years, these logbook entries will accumulate almost all the aspects of the instruments in the observatory. Although the key technologies are changing rapidly, the fundamental principles involved in construction and maintenance at observatories are getting altered at a less rapid rate. This fact necessitates the need for keeping a record of activities carried out over the years for prompt diagnostics. Projects like SKA, TMT, LIGO, SALT, JWST also require extensive internal coordination. These typically are a collaboration consisting of thousands of scientists whose research can span area like instrument fabrication, installation, commissioning, characterization, maintenance, upgrade, data analysis and parameter estimation. Often their time span spread across few decades and thus generate information whose volume and complexity cannot be handled effectively by traditional search engine backed information processing tools. On the positive side, analysis of such big data volumes can yield powerful insights into the inherent trends and fluctuations within the concerned project. 

 In this paper, we demonstrate natural language processing(NLP) backed knowledge rediscovery \citep{ricci2011introduction} by making use of the open source logbook data from the Laser Interferometric Gravitational Observatory (LIGO).  This is a novel approach to observational astronomy, and  the developed software is made available for the public through a web application named \textbf{Hey LIGO\footnote{\url{heyligo.gw.iucaa.in}}}. We also show the application of descriptive content based recommendations to compare common issues among multiple observatories. These methods are generally scalable and will be very useful in the event of upcoming projects like the Square Kilometer Array (SKA) and upcoming LIGO-India detector.

We have organized the paper as follows: in section two we describe the methodology adopted to convert raw data to useful and representable information. Section three provides the details of data used in our analysis. Features of the recommendation system are outlined in section four. In the last part, we apply the scheme to various gravitational wave observatories around the world and discuss the results obtained.

\section{Contextual learning of unstructured Data}
Structured data is highly organized and usually resides in a relational database schema. But unstructured data refer to information that does not follow the traditional database scheme.  For example, e-mails, web pages, business documents, FAQ's, etc. are some examples of unstructured data. They include text and even multimedia content. So processing of such information is an energy and time consuming task. 

\begin{figure*}[!htb]
\begin{center}
\includegraphics[scale=0.4]{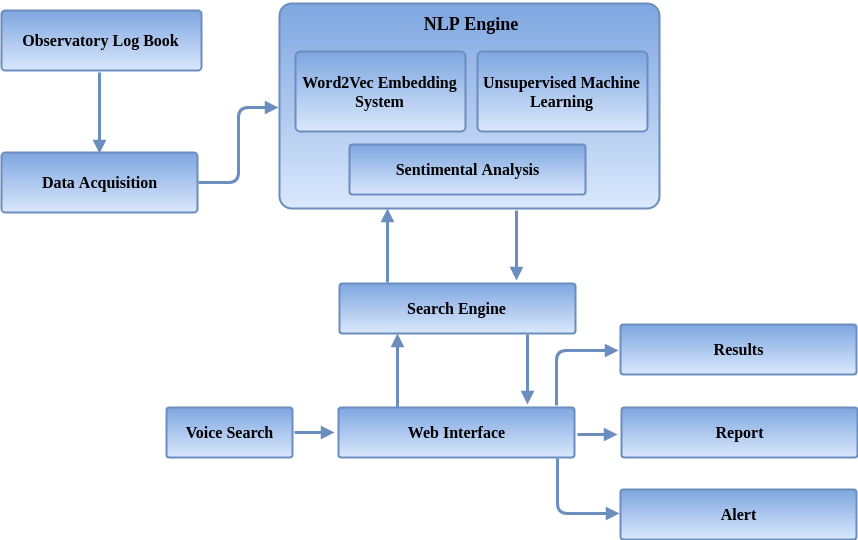}
\caption{Data Flow through various components in the system}
\label{fig:flowchart}
\end{center}
\end{figure*}

This section briefly describes the manner in which the unstructured textual data is acquired, processed and finally given structure. Moreover, it also enumerates the various steps involved in the development of a machine learning model which is used to differentiate between the available textual data points contextually. Finally, the model is used to perform clustering over all the textual data, thus, adding a structure for ease of access. Figure~\ref{fig:flowchart} shows the schematic representation of our web interface that is used to implement the scheme.

The unstructured dataset that we use is in the form of textual web pages. These pages have an identical HTML structure and defined attributes for every data point. Due to this same structure and open source nature of the web pages, it is possible to write a script which can extract each attribute from the HTML source code and organize the complete information into a data frame. A data frame is a tabular structure with columns as attributes and rows as individual data points. The first part of our algorithm does data acquisition using the python package "Beautiful Soup" to retrieve information from web pages by searching through all the new posts and related data. It saves it into relevant files for later  utilization when the need arises.

Once the data is stored locally, non-essential attributes are removed, textual time stamps are converted to system timestamps, duplicate data points are removed/combined, and the resulting data is passed on to the text processing unit. A vocabulary for our data is generated by converting the unstructured data into stem words. For that, we have removed all special characters and punctuations such as $!, @, \#, \$, \%, * ,\&, ',",(,) $ etc.  All non-English words and other HTML tags, URLs are also excluded from the data.  The text is then tokenized \citep{huang2007rethinking} by splitting the strings of text into a list of words called tokens.  To reduce the redundancy in the vocabulary, it is important to convert the related token forms and their derivatives to a common base stem by a process known as stemming  \citep{smirnov2008overview}. 

In the process of embedding, the textual data is converted into vectors which can be easily handled by the computer \citep{li2015word}.  There exist various embedding algorithms like One-Hot Encoding \citep{harris2012digital} and Term Frequency Inverse Document Frequency (TFIDF)  \cite{leskovec2014mining} etc for this process. Not all methods can capture the contextual differences between the words. However, a recent breakthrough in the field of Natural Language Processing incorporates neural networks that can learn the vector values for each word by iterating over the text multiple times using a gradient based algorithm \citep{2013arXiv1301.3781M,2013arXiv1310.4546M}. \cite{bengio2003neural} have coined the term word embeddings with a neural language model to train them with the model's parameter.

One of the commonly used tools to convert words into vectors is word2vec described in  \citet{2013arXiv1301.3781M}. Word2vec has a single hidden layer, fully connected neural network that takes a large text corpus as input and produces a higher dimensional vector for each unique word in the corpus.  Words which share common contexts in the corpus are located close to each other in the vector space. Word2vec models do not consider word order and can capture semantic information between words in a very efficient way \citep{ling2015two}. With the help of Word2vec embeddings, a computer can differentiate between words of different types. Word2vec implements two computationally less expensive models known as Continuous Bag of Words (CBOW) and a Skip Gram model \citep{2013arXiv1301.3781M} to learn word embeddings. The representation of a corpus of text or an entire document in the form of a list of words (Multiset) is referred to as Bag of Words representation \citep{markov2007data}.  The algorithm essentially tries to predict the target based on a set of context words \citep{2013arXiv1301.3781M,2013arXiv1310.4546M}. 

\begin{figure}[!htb]
\begin{center}
\includegraphics[scale=0.25]{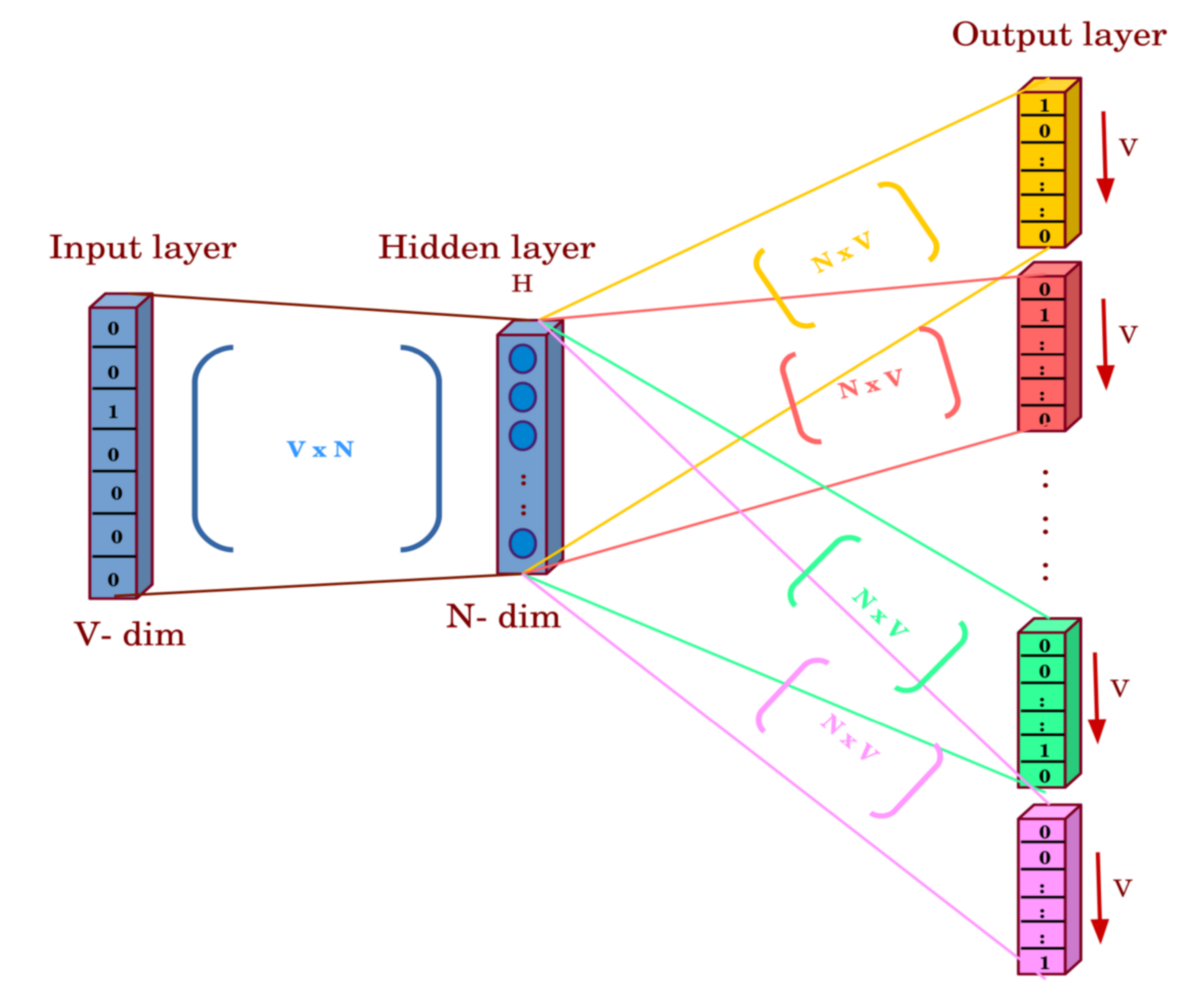}
\caption{Skip-Gram Model}
\label{fig:skipgm}
\end{center}
\end{figure}

The model that we have used in this work is the Skip Gram model. The basic architecture of the skip gram model is shown in Figure~\ref{fig:skipgm}.  This representation is similar to CBOW model, but instead of predicting the target word, it predicts the context words based on a given target word \citep{2013arXiv1310.4546M}. Thus, the model maximizes the probability for classification of a word based on another word in the same sentence \citep{2013arXiv1301.3781M}.  Thus the vector representation is capable of capturing the semantic meanings of the words from a sequence of training words $w1, w2, ..., wT$ and their contexts $c$.  The algorithm can be briefed as follows. First the words are applied to as an input to a log-linear classifier where the objective is to maximize the average log probability given by,

\begin{equation}\label{eq:skpg}
\mathcal{L} = \frac{1}{T}\sum_{t=1}^{T} \sum_{-c\leq j \leq c,j\not\equiv 0} log \ P(w_{t+j} | wt) 
\end{equation}
Larger value of $c$ can result in higher accuracy but requires more training time \citep{2013arXiv1310.4546M}. To obtain the output probability, $\mathit{P(w_{o}|w_{i})}$, the model estimates a matrix which maps the embeddings into $V$-dimensional vector $O_{w_{i}}$. Thus the probability of predicting the word $w_{o}$ given the word $w_{i}$ is defined using the softmax function:

\begin{equation*}
\mathit{P(w_{o}|w_{i})} = \frac{\exp{(O_{w_{i}}(w_{o}))}}{\sum_{w \epsilon \mathit{V}} \exp{(O_{w_{i}}(w))}}
\end{equation*}

where $V$ is the number of words in the vocabulary \citep{ling2015two,2013arXiv1310.4546M}. But this formulation is computationally intensive for larger vocabularies. This is solved in word2vec by using the hierarchical softmax function \citep{morin2005hierarchical} or with negative sampling approach \citep[see this for more details;][]{2014arXiv1402.3722G}.

After embedding all words,  every data point is represented as the average of all the word vectors of the words present in it. A Nearest Neighbor Algorithm \citep{Andoni:2006:NHA:1170136.1170526} is then used to cluster these data point vectors to respective clusters efficiently.  The optimal number of clusters is estimated iteratively until it is observed that the accuracy peaked, which in our case was found to be $1/5^{th}$ of the vocabulary of our model. We used the Python implementation of $Scikit-learn$ package for doing the nearest neighbor algorithm.

Even after NLP classification, we observed that quite a number of relevant posts were left out unobserved and so we added one more layer of processing  by analyzing the overall emotional content of the reports. We used the AFINN lexicon \citep{AFINN} consisting of a collection of 2477 words each with an associated integer value ranging between -5 to +5 representing transition from negative to positive sentiment. Modifying the word valence and appending the lexicon with technical words that better represent the associated sentiments was found to  provide better results. For example, LIGO specific application will associate terms like 'lockloss' and 'scatter noise' with negative sentiment while 'new filter installed' would mean something positive.

\begin{figure*}[!htb]
\begin{center}
\includegraphics[width=16cm]{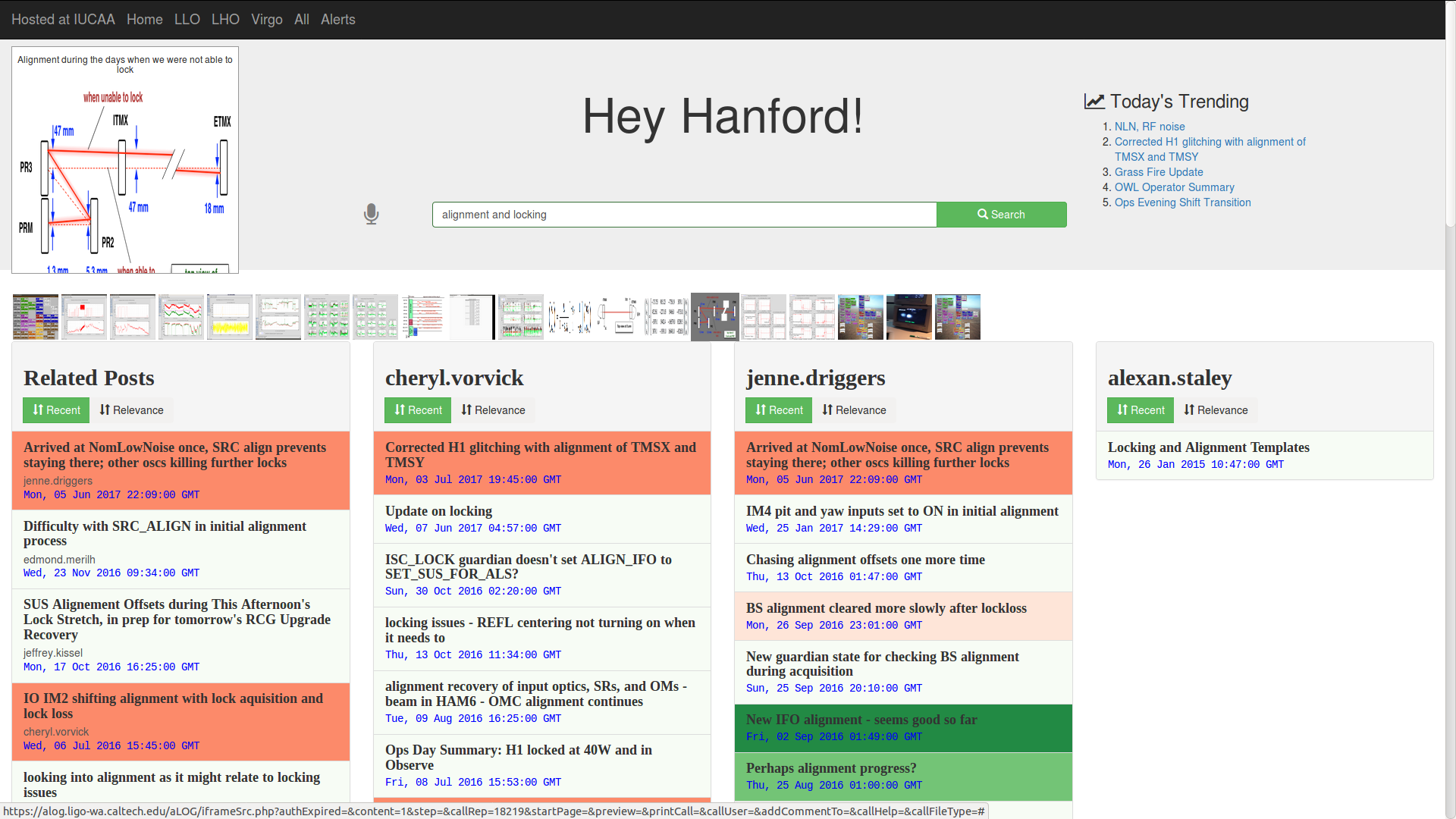}
\caption{Screen-shot of Hey LIGO Web Interface. Search results are color tagged based on their overall sentiment. Trending posts are identified based on the associated meta-data consisting of comments \& discussions made within the LIGO community }\label{fig:respg}
\end{center}
\end{figure*}

The prototype is designed so that the users can query for information through a web interface.  The stem words in the query are identified and the vector is projected into the previously modeled word vector space.  The nearest neighbor model retrieves the top neighbors for the query vector, and are then displayed as the search results on the web interface. The search result is then filtered to check for the presence of the query words in either the title  or in its content to weed out false positives. In Figure~\ref{fig:respg}, we have shown a simple search query displayed on the web interface. The different features incorporated in the web page is described in Section~\ref{sec:HFunc}.

\section{Gravitational Wave Observatories }\label{sec:Test}

GW interferometers (IFOs) have been in operation for the past few decades and have made the first direct detection of merging binary black holes \citep{PhysRevLett.116.061102}. The complex nature of this multi-physics experiment requires  scientists from multiple domain of expertise to work together and share information. Rigorous commissioning and characterization efforts have been carried over a span of two decades to reach the current level of sensitivity. LIGO, Virgo, GEO600 and KAGRA  archives most of the activities happening at the sites through their logbooks. These may be complete from installation activities to noise hunting and mitigation works carried out during the lifespan of the observatory. Although there are site specific issues, often they encounter problems of similar nature and employing solutions that worked at the other sites may be a good strategy to start with. Also it is not uncommon to see previously fixed issues to reappear at a later time where the time scales could be of few months to years. This happens due to recurring environmental fluctuations and configuration changes in the detector. Since the current GW detectors aim at coincident detection of events, joint uptime of the instruments are crucial. This is more significant because the probability of detection scales linearly with observation time and cubically with sensitivity of the instrument.  

The effective information extraction and processing of logbook information as envisaged here is expected to help in making better decisions pertaining to detector maintenance. For example, identifying the subsystems that could possibly get affected during instrument upgrade will be vital in scheduling and coordinating tasks among sub-groups involved. Similarly, long term tracking of a issue can be carried out to see if the various overhauling attempts indeed lead to an improvement in performance which correlates with lesser number of related posts.

In the case of GW interferometers, day-to-day activities are recorded using web interfaces known as Logbooks. It is mandatory for the reports to have a title, section, task, details and author details. Although anyone can view the reports, only users with valid credentials can login to add logs and additional supporting files like measurement figures, sensor data, codes etc. It is also possible to to add comments and carry out further discussion on any of the log book entries. Details of retrieved information are given below in Table~\ref{tab:logs}

\begin{table*}[!htb]
\begin{center}
\caption{Log book details retrieved from different GW observatories.} \label{tab:logs}
\begin{tabular}{|l|c|c|c|c|c|}
	\hline
	Observatory & Logbook Entries & Contributors & Timespan & Dictionary Size & Clusters\\
	\hline
    LIGO Livingston\footnote{\url{https://alog.ligo-la.caltech.edu/aLOG/}} & 24351 & 261 & 2010-2017 & 2273 & 455\\
	\hline
    LIGO Hanford\footnote{\url{https://alog.ligo-wa.caltech.edu/aLOG/}} & 24968 & 237 & 2010-2017 & 2713 & 543\\
    \hline
    Virgo\footnote{\url{https://logbook.virgo-gw.eu/virgo/}} & 34592 & 660 & 2010-2017 & 5026 & 1005\\
	\hline
\end{tabular}
\newline

\end{center}
\end{table*}

\section{Hey LIGO Functionalities}\label{sec:HFunc}

An open access NLP based web application implementation named \textbf{Hey LIGO} is developed and deployed to support the commissioning and characterization efforts at the GW observatories. It relies on the logbook data recorded since 2010 by scientists specialized in different aspects of the detector. Every query is answered by matching it with most relevant logbook entries sorted as per their closeness to the query term in the word-vector space. We further analyze the sentiment of the post and color code so that green indicates a positive outcome and red corresponds to something undesirable in the context of activities carried out at the detector. An image retrieval facility displays thumbnail of the figures attached to the sorted data simplifying the knowledge discovery process.  Contextual data visualization across multiple detectors is carried out as shown in Fig \ref{timeline} and Fig \ref{scatter_pie}. This feature lets the user to compare and see the trends in the searched keyword across different observatories. 

Automatic check for new data entries is done periodically so that the NLP models are regularly  updated. We track the volume of discussions happening on various topics and hence identify and rank the trending issues on a daily basis. Scientists involved with the project will mostly be interested in getting notified about specific issues that correlates with their domain of expertize and so the application only issues alerts to registered participants with matching interests. This targeted delivery will remove clutter and will ensure proper dissemination of information to the concerned people. 

Code development is usually a tedious procedure wherein significant amount of time is spent on readability and re-usability so as to benefit a wider research community. Our application makes better use of this idea by auto-detecting and notifying the user about the presence of codes in the searched content. This feature we believe would simplify the procedure involved in result reproduction and its consequent independent verification.

To check the performance of our application, we analyzed six months of logbook data from LIGO Livingston and compared the NLP results and the actual relevant entries. Table~\ref{tab:acur} gives the performance of the our implementation for certain set of randomly chosen keywords. In most cases,the false alarms occur at the tail end of the search results which represent neighbors of least relevance in the k-NN search. These can be removed either by setting a threshold on the similarity distance measure or by post-filtering the results by additionally comparing the content of each post. Currently, we have implemented the process of post-filtering to remove such post in the final web application. In the future we are planning to incorporate a mechanism that will make use of feedback received from the users and utilize it to improve the accuracy in retrieving relevant posts.  

\begin{table}[!htb]
\begin{small}
\begin{center}
\caption{Table showing some of the alog posts from May $1^{st}$ to June $30^{th}$, 2017.} \label{tab:acur}
\begin{tabular}{|m{5em}|m{1.2cm}|m{1.2cm}|m{1.2cm}|m{1.2cm}|}
	\hline
    Keyword & {Log book Entries} & \multicolumn{3}{c|}{Posts retrieved by NLP code} \\
    \cline{3-5}
    &  & \scriptsize{Total}  & \scriptsize{Relevant} & \scriptsize{Irrelevant}\\
	\hline
    Lock loss & 108 & 108 & 89  & 19\\
	\hline
    Earthquake & 83 & 94 & 80  & 14\\
    \hline
    \scriptsize{Charge measurement} & 62 & 65 & 58 &  7\\
    \hline
    Guardian & 55 & 65 & 55  & 10\\
    \hline
    Oplev & 63 & 61 & 48 & 13\\
    \hline
    \scriptsize{Calibration lines} & 55 & 52 & 45 & 7\\
	\hline
\end{tabular}
\end{center}
\end{small}
\end{table}

\section{ Inferring from logbook entries}
Once the relevant logbook entries are identified using the techniques mentioned above, their associated meta-data can be utilized to obtain several quantitative information about the topic of interest. 
\pagebreak
\subsection{Trends within detectors}

Below we briefly compare the trends obtained for few test search queries and briefly discuss the observed patterns. Although of similar configuration, the effect of various noises on each detector seems to be of a different nature.  Variation in instrumental behavior and environmental effects due to geographical location will also influence the efficiency of implemented mitigation measures. Figuring out such details can positively speed up the commissioning activities of future detectors like LIGO-India. 

\begin{figure*}[!htb]
\begin{center}
\includegraphics[width=14cm]{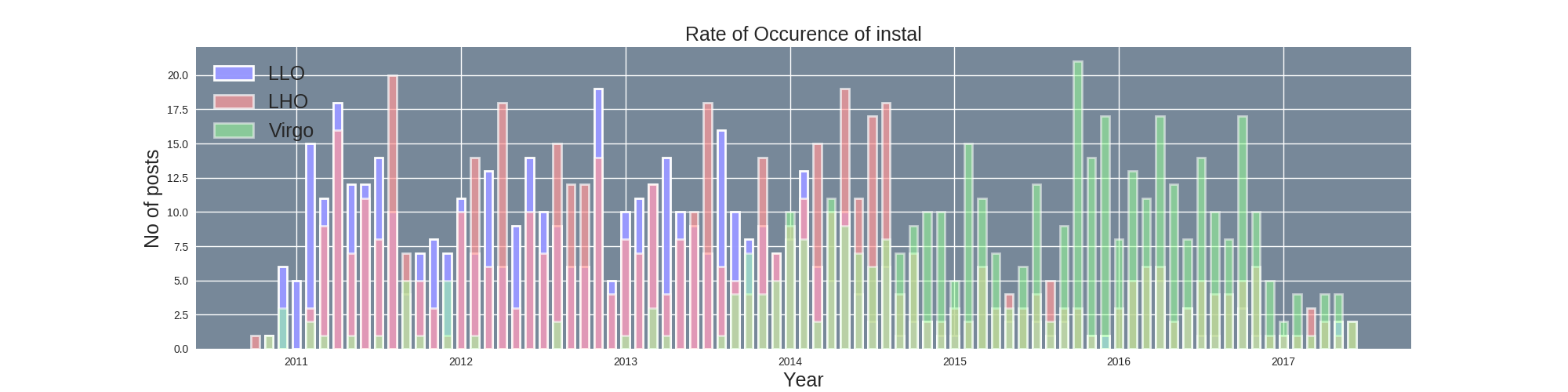}
\includegraphics[width=14cm]{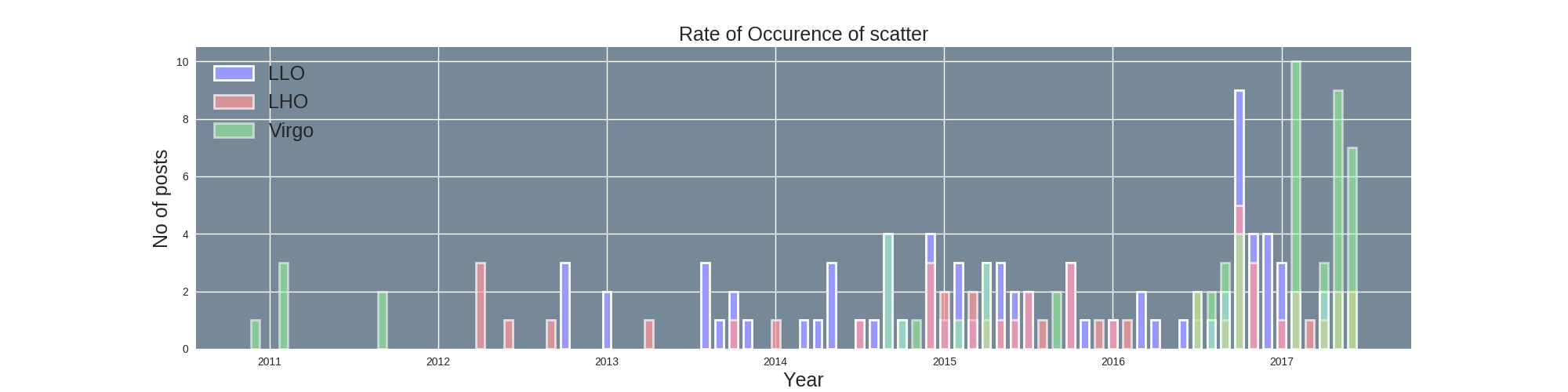}
\includegraphics[width=14cm]{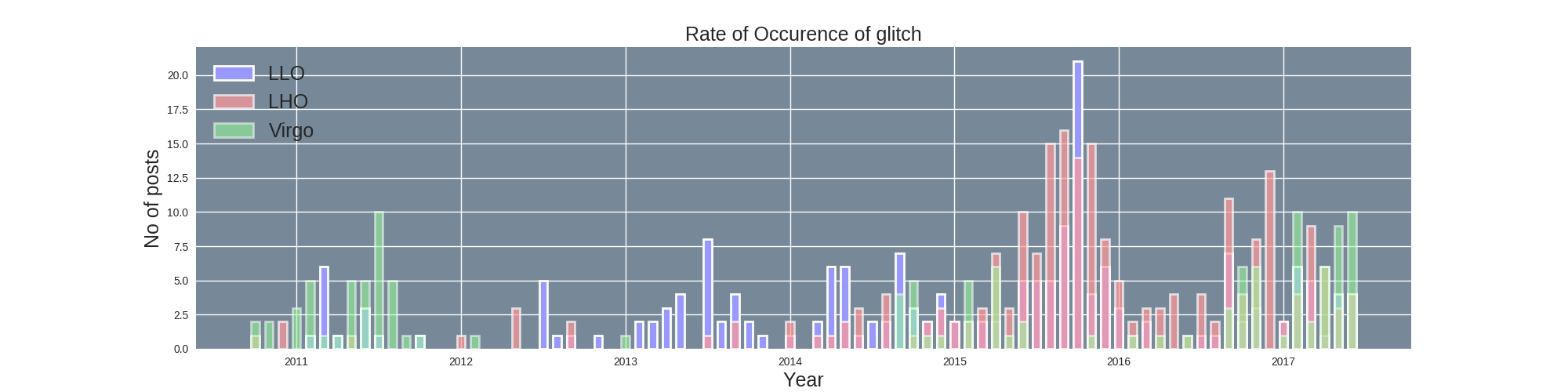}
\includegraphics[width=14cm]{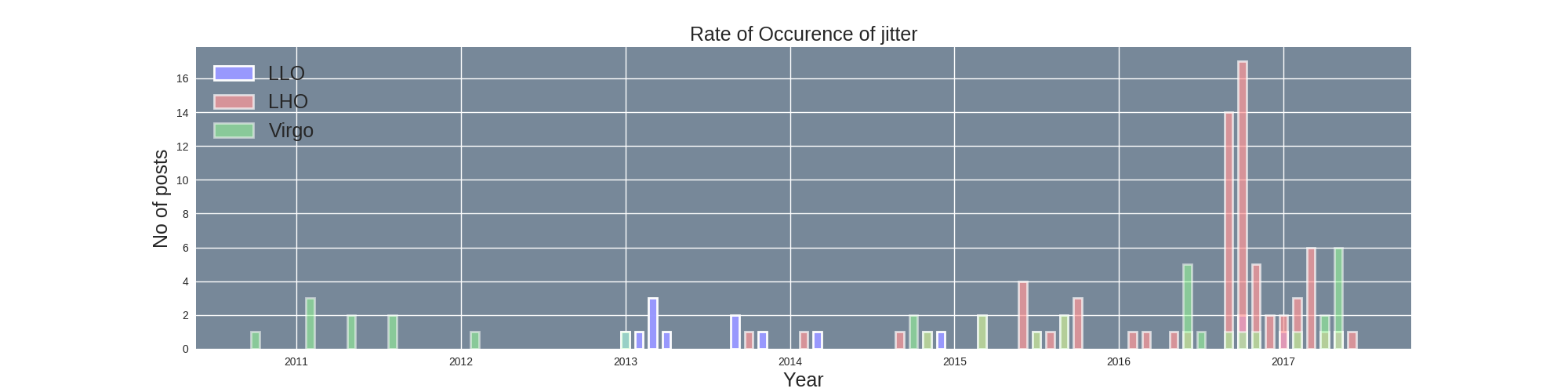}
\caption{ Rate of occurrence of different key words in multiple detectors are a function of time}
\label{timeline}
\end{center}
\end{figure*}

\begin{enumerate}

\item \textbf{ \small{Installation}}

First plot from figure \ref{timeline} shows the trends in posts related to installation work at each of the observatories. Activities picked up momentum in 2010 at LIGO and continued till the mid of 2014 after which testing and commissioning tasks started. Advanced Virgo seems to have started such activities in 2014 and carried on till the end of 2016. 

\item \textbf{\small{Jitter Noise}}

Jitter noise arising out of laser pointing fluctuations \citep{PhysRevD.93.112004}, is sensitive to cavity alignments and angular mirror motions. It has been partly caused by the pre-stabilized laser (PSL) periscope motion induced by chiller water flow around PSL's high power oscillator. Various efforts to understand it's possible origin and subsequent efforts to subtract it from the data stream is reflected through the increased number of alogs at the Hanford detector compared to other sites. Commissioners performed online feed-forward noise subtraction using auxiliary witness channels which reduced the coupling significantly (\citeauthor{alogLLO30412, alogLLO30473,alogLLO34631}).

\item \textbf{\small{Scattering Noise in LHO and LLO}}
Noise from scattered light is currently one of the factors that limit the sensitivity in the frequency bin from 50 Hz to 200 Hz (see Fig.~\ref{scatter_modelled}) especially during periods of high microseism. Off-axis beam scattered laser beam could hit a reflecting surface like camera mirror mount or beam tube and reenter the cavity. Nonlinear features are seen in the gravitational wave spectrum when this beam picks up resonances from reflecting surfaces which then get upconverted or phase modulated by low-frequency seismic-like motion. It's effect at LLO is more pronounced as compared to LHO as the former is vulnerable to microseismic activity \citep{Ottaway:12}.  

\begin{figure}[!htb]
\begin{center}
\includegraphics[width=10cm]{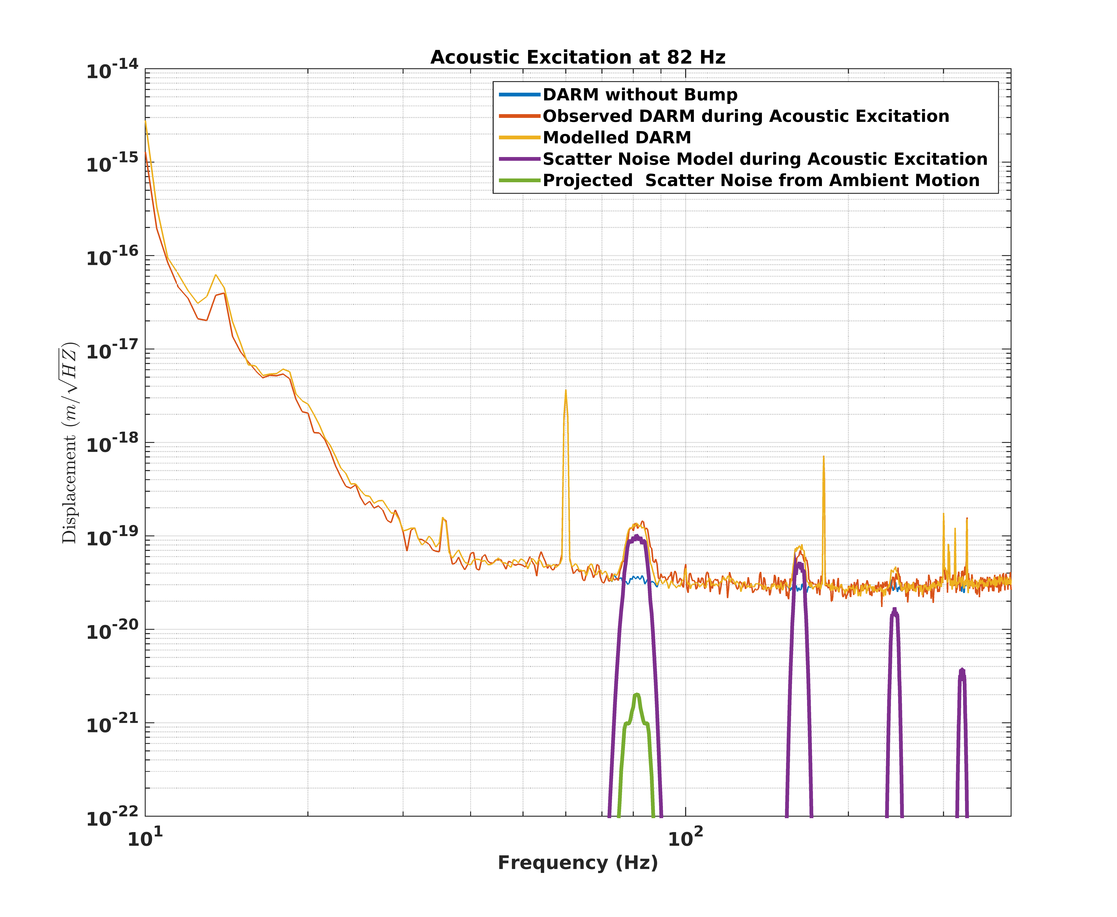}
\caption{ Noise due to scattered light observed at LIGO Livingston. Noise gets amplified and upconverted during periods of high microseism and limits the sensitivity range of the GW detectors. }
\label{scatter_modelled}
\end{center}
\end{figure}

Figure \ref{scatter_modelled} shows the effect of acoustic excitation on  82 Hz peak seen in gravitational wave differential arm motion (DARM) data.  The acoustic injections carried out at LIGO Y-end station are reconstructed using the proposed model  \citep{VirgoScatterPaper}  Scatter Noise = A sin(4*pi* (n*Yrms + Yac)/lambda) where Yac = B*sin(2pi fo t) where Yrms is the ground motion and Yac is the chamber motion with (A,n,B) being the tunable parameters. Model parameters using are fine tuned using pattern search. The scatter noise projection to DARM from ambient motion is obtained by scaling down the chamber motion based on the accelerometer signal before and after injection.

\begin{figure*}[!htb]
\begin{center}
\includegraphics[width=8cm,trim={1.2cm 1.2cm 1.2cm 1.2cm},clip=true]{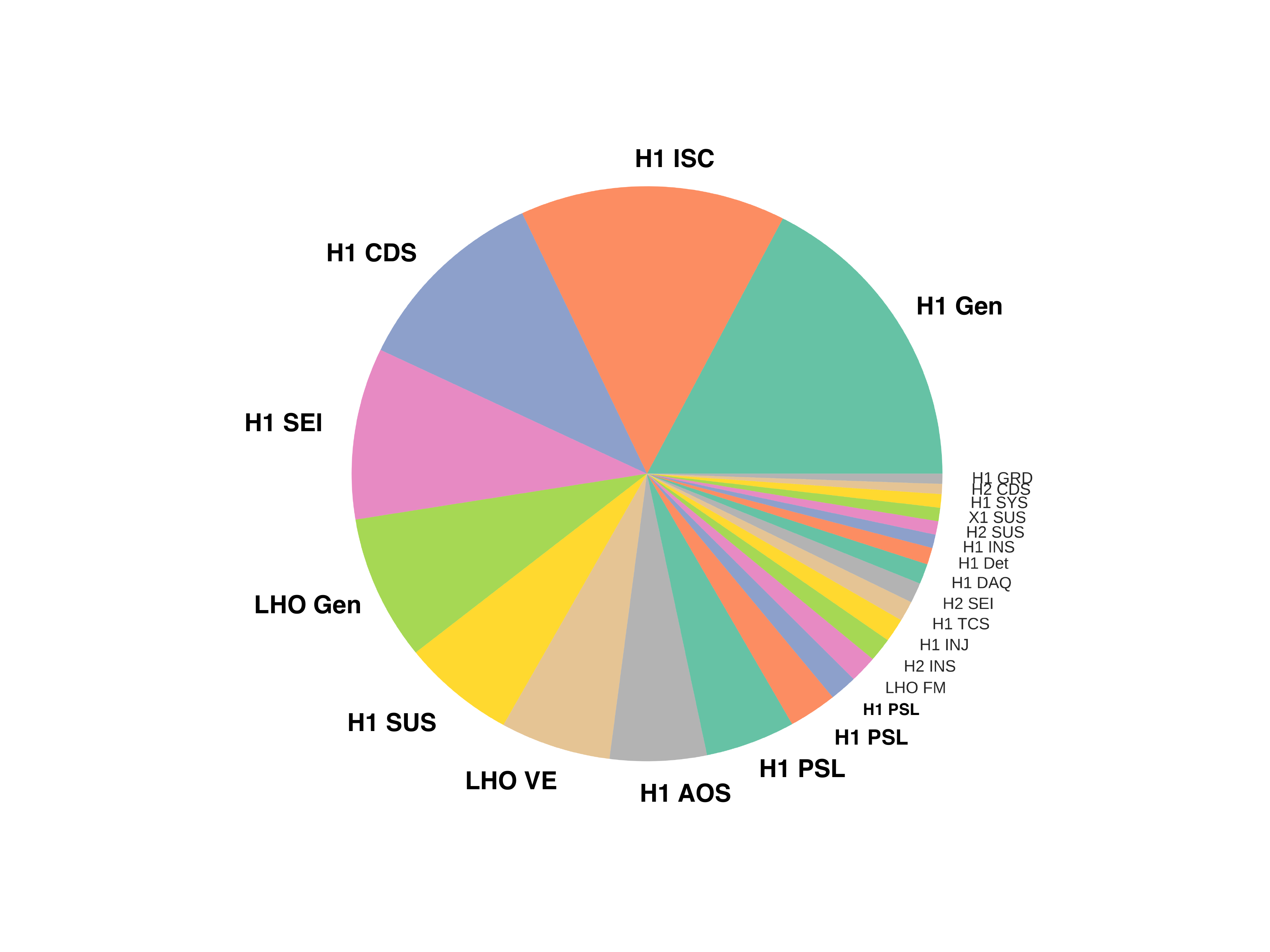}
\includegraphics[width=8cm, trim={1.2cm .21cm 1.2cm 1.2cm},clip=true]{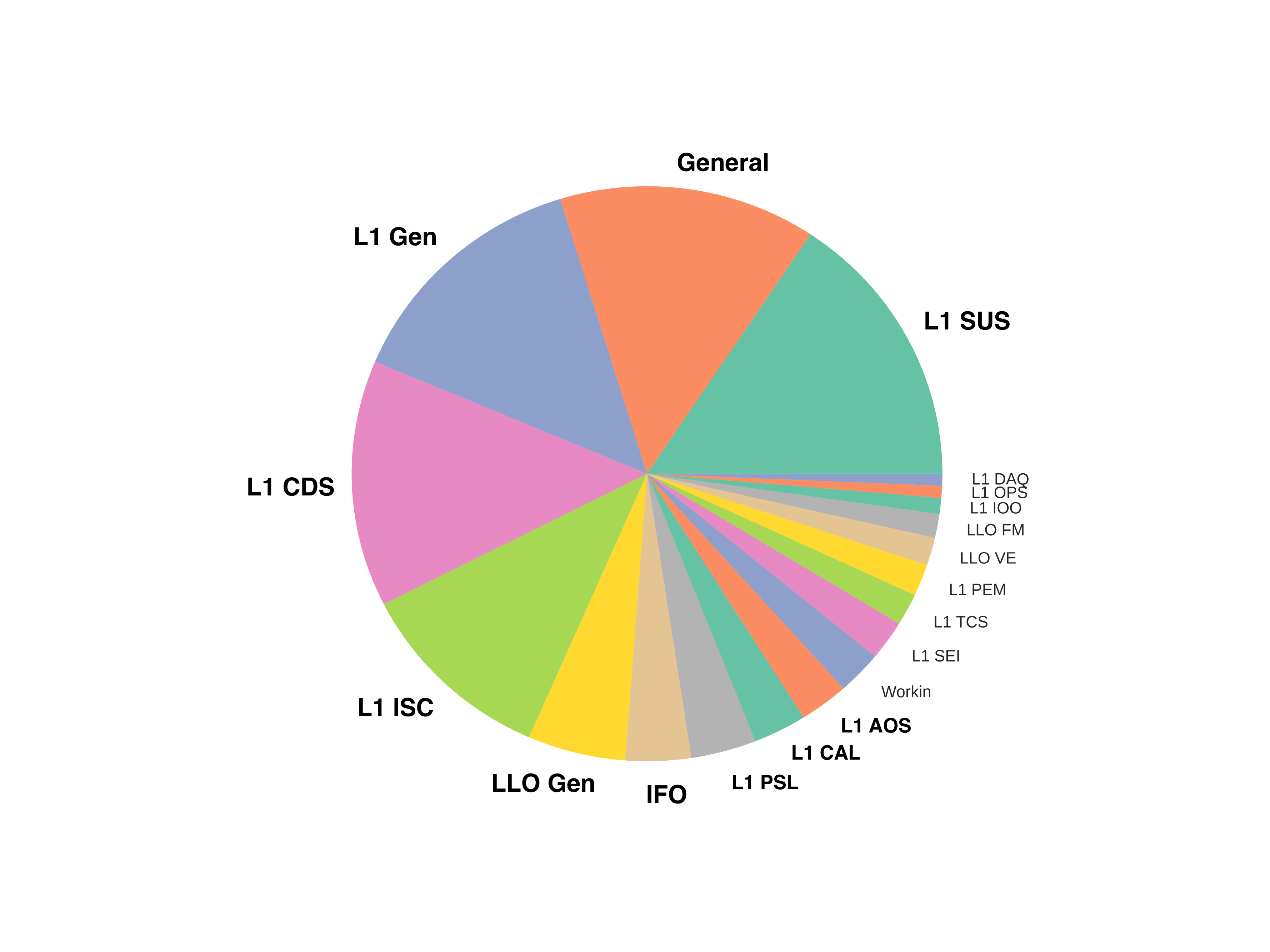}
\caption{\small{ Pie Diagram shows how various LIGO subsystems contribute to the various non-astrophysical glitches seen in LIGO. Acronyms Used, \textbf{ISC}: Interferometer Sensing and Controls, \textbf{CAL}: Calibration, \textbf{AOS}: Auxiliary Optics Support, \textbf{SUS}: Suspension, \textbf{VE}: Vacuum Electronics, \textbf{SEI}: Seismic External Isolation, \textbf{CDS}: Control and Design System } }
\label{scatter_pie}
\end{center}
\end{figure*}

\item \textbf{ \small{Non-astrophysical Transients }}

Glitches often show up in the strain data leading to false alarms in the various search pipelines that look for astrophysical signals. Triggers are also observed in badly functioning instruments and are witnessed in auxiliary sensor channels. Some of them are also reported to cause loss of lock of the interferometer. The general operation of all the three detectors have been affected by such transients ever since their beginning of operation \ref{timeline}. The report generation feature of our application provides the following glitch distribution (Fig. \ref{scatter_pie}) across multiple subsystems based on their tags in the data. It is interesting to note the subtle variations in the noise sources between LHO and LLO. The origin of many of them have been studied and reported in the logbooks while a vast majority are still not well understood. 

\end{enumerate}

\subsection{Visualizing Observatory as a Complex Network}

Behavior of an observatory and the elements that lead to changes within the system behavior can be studied through its representation as a complex network. Complexity is expressed through nodes and links within the network. 
Here, the nodes can be either subsystems or specific instruments or even subgroups within the observatory and edges between them provide the probability of each one them being connected to the other as inferred from logbook entries. We first create a dictionary of subsystem keywords and for each one, find the frequency of their joint occurrence with everyone else. This information is then used to form the adjacency matrix whose diagonal elements are all zero and the off-diagonal value representing the linkage is given by the ratio of joint occurrence frequency divided by total occurrence of the keyword. Adjacency matrix being non-symmetric leads to a directed graph. Number of incident edges determines the node size while the edge width is given by the associated connection probability. To better aid visualization, we adopt Force Atlas 2 layout \citep{jacomy2014forceatlas2} with repulsion being approximated using Barnes Hut optimization \citep{barnes1986hierarchical} which is well suited for larger graphs. The interconnectedness information within the observatory revealed through these networks may help in identifying the critical nodes in the system and the makes it easier to identify the vulnerable connections. These representations could possibly be useful during large scale repair and maintenance as  they reveal the other subsystems that can get affected in the process. 

\begin{figure*}
\begin{center}
\includegraphics[width=14cm]{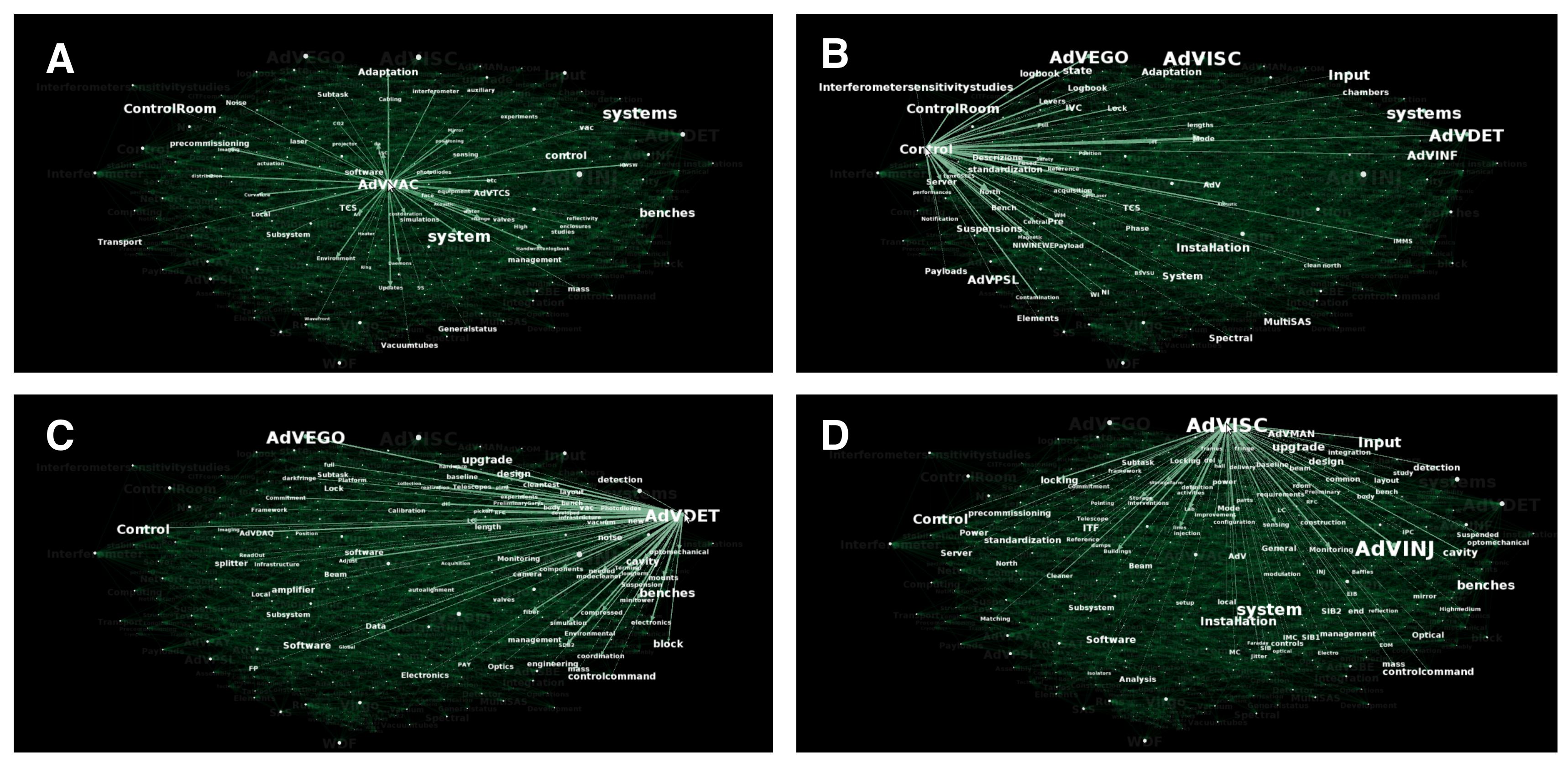}
\caption{Network plot of the Virgo detector highlighting the inter-connections between various subsystems. Subplots highlight the directed graph for A) Vacuum, B) Control system, C) Photodetector and D) Interferometric sensing \& control.}
\label{VirgoCombo}
\end{center}
\end{figure*}

In Fig ~\ref{VirgoCombo} we show the network connection for a few prominent nodes of Virgo observatory. It differs from real-world networks in terms of its degree distribution (degree refers to the number of edges connected to each node). Sparse networks are characterized by a degree distribution which takes form of a power law and are commonly seen in biological networks and  computer networks. \citep{barabasi2016network}.  For the case of Virgo network, this distribution deviates from such a power law indicating dense connection between the nodes. Further research is needed to analyze the network and study the instrument's robustness to random sub-system failures.

\section{Discussions \& Conclusion}
We have demonstrated how information retrieval and recommendation systems could be useful for LIGO like astronomical observatories. Compared with conventional search associated with the existing sites, our web application incorporates a natural language processing based information retrieval system which can also do visualization of the user queried data. Involving a wider science community in big science projects can alleviate some of the issues related to lack of sufficient man power within the project. The developed interface identifies the major issues based on the discussions done within LIGO community and recognizes the trending issues. It is plausible that someone outside the project has already seen and solved these before. Hence  proper dissemination of information will help in technical experts outside the project collaboration to contribute improving the overall performance of the instrument.

Coordinated efforts are being undertaken worldwide to carry out electromagnetic follow-up searches looking for counter parts to coalescing binaries sources\citep{abbott2016localization}. During the instance of GW candidate event alert, astronomers may be able to take advantage of our application and know more about the instrument. 

Future improvement in the application would be to include  capabilities wherein an identified issue will be provided with possible fixes making use past attempts which fixed an identical issue. This would require text abstraction and summarization, quite challenging when the data has ample amount of technical terms. Efforts to add other GW detectors like GEO600 and KAGRA is currently under progress and will enhance the effectiveness of our application.

This kind of system has a lot of potential applications with the commissioning and running of large science projects like the SKA and future LIGO observatories. In this project the data source was more unstructured and had few tags related to the status of different activities. At present institutions like SKA South Africa which is in charge of building MeerKAT \footnote{\url{http://www.ska.ac.za/science-engineering/meerkat/}} telescope which is one of the precursors to SKA, use a more structured systems like JIRA \footnote{\url{https://www.atlassian.com/software/jira}} for issue tracking and log keeping\footnote{\url{https://indico.skatelescope.org/event/402/material/1/6.pdf}}. Scaling our present system to such databases can improve the efficiency of topical modeling. This also enables auto-update of the learning database as more and more information is logged into the system finally making it a robust. 

The aforementioned feature will eventually be very useful when the organization grows with the number of participants increasing over time. Availability of such systems will make the re-usability of information much easier and efficient. This will also help in resolving instrument issues much easier and faster. Enhanced analytics of key components and recurring issues can help improving the fault tolerance of different subsystems and could provide insights on how modify them for better performance.

\section{Acknowledgements}

We would like to thank  detector characterization group and machine learning sub-group of the LIGO Scientific Collaboration for their comments and suggestions. NM acknowledges the Council for Scientific and Industrial Research (CSIR), India for providing financial support as a Senior Research Fellow and Navajbai Ratan Tata Trust (NRTT) grant for supporting his visit to LIGO Livingston. A. K. Aniyan would like to thank the SKA South Africa postgraduate bursary program. SM acknowledges the support of the Science and Engineering Research Board (SERB), India through the fast track Grant No. SR/FTP/PS-030/2012. Authors express thanks to  Arnaud Pele, Anamaria Effler and Ajit K Kembhavi for their valuable comments and suggestions. Authors wish to thank Malathi Deenadayalan and Santosh Jagade for technical support. LIGO was constructed by the California Institute of Technology and Massachusetts Institute of Technology with funding from the National Science Foundation and operates under Cooperative Agreement No.PHY-0757058. This paper has been assigned LIGO Document No. LIGO-P1700250.


\bibliography{Mukund_HeyLIGO_2017}



\end{document}